\documentstyle[amstex,aps,amsfonts,prl]{revtex}

\begin{document}
\title{Dynamics of classical wave scattering by small obstacles}
\author{Gerrit E.W. Bauer,$^{1}$ Mauro S. Ferreira,$^{1}$ and Cees P.A. Wapenaar$%
^{2} $}
\address{$^{1}$Theoretical Physics Group, Department of Applied Physics and \\
Delft Institute of Microelectronics and Submicron Technology, \\
Delft University of Technology, \\
Lorentzweg 1, 2628 CJ Delft, The Netherlands\\
$^{2}$Department of Applied Earth Sciences, Delft University of Technology, 
\\
Mijnbouwstraat 128, 2628 RX Delft, The Netherlands}
\date{\today}
\maketitle

\begin{abstract}
A causality problem in the time-dependent scattering of classical waves from
point scatterers is pointed out and analyzed. Based on an alternative model,
the leading pole approximation of the exact scattering matrix of the square
well potential, transparent expressions for the time and position dependent
Green function in a disordered medium are derived.
\end{abstract}

\pacs{PACS numbers: 42.25.Dd, 46.40.Cd, 46.65.+g, 91.30.Fn }

Wave propagation in complex media is a large and interdisciplinary field of
research with many unsolved problems that are scientifically challenging and
technologically important \cite{general}. Electromagnetic and acoustic waves
are massless and described by a ``classical'' wave equation which is second
order in the time derivative, whereas ``quantum'' particle waves are
governed by the Schr\"{o}dinger equation which is first order in $\partial
/\partial t$. Scalar classical waves at a fixed frequency $\omega $ are thus
equivalent to a particle wave at energy $E\sim \omega ^{2}$. The analogies
between the classical and quantum problems indeed lead to many cross
fertilizations since solutions obtained in one field can be carried over to
the other. For example, the discovery of localization of light in random
media \cite{Wiersma} was stimulated by earlier work on electron localization 
\cite{Lee}. Disordered systems have to be treated by statistical methods,
for example by an ensemble average over the configurations of randomly
distributed model scatterers. The (diagrammatic) perturbation theory for
potential disorder in metals \cite{Abrikosov} has been helpful in
understanding classical wave propagation, including (weak) wave localization 
\cite{Rossum}. An essential ingredient in these calculations is a
mathematically simple yet physically meaningful and well-behaved model for
the disorder. The point ($\delta $-function) scatterer is often the basic
building block for classical and quantum problems, because of its simple
scattering amplitude. The associated ultraviolet divergence must be
regularized, however \cite{Vries,Rossum}.

Most previous studies of classical waves in complex systems have been
limited to monochromatic, steady-state wave fields, but the propagation of
short pulses containing a broad band of frequencies is of considerable
interest as well. Concrete challenges are provided by seismic waves excited
by earthquakes or artificial explosions in terms of first arrival times \cite
{Snieder} and the waves which trail behind (the so called ``coda'') \cite
{Coda}. The power spectrum of the coda has received theoretical attention 
\cite{White}, also in the framework of random matrix theory \cite{Titov}.
Random matrix theory of transport, up to now mainly applied to quantum
problems \cite{Beenakker}, identifies the ``geometrical'' or ``universal''
features of an observable that do not depend on the microscopic details of
the system. Alternatively, however, one may pursue a microscopic approach
trying to answer, for instance, what sort of information about the
geophysical complexity of earth subsurface can be distilled from the {\em %
total} seismic trace. To this end models and approximations must be found
which are well-behaved and sufficiently accurate over the whole complex
frequency plane. In this Letter we point out that point defects interacting
with classical waves are responsible for wrong analytical properties of the
scattering matrix in the frequency domain equivalent to non-causality in the
time domain, which appears to have escaped attention in the literature. We
propose a different model that has all practical advantages of point
scatterers but gives a causal response with the additional bonus that an
ultraviolet cutoff is not needed. For simplicity, we limit attention in the
following to scalar fields referring to \cite{Vries} for a discussion of
point scatterers for vector fields. After treating the simple
one-dimensional (1D) problem, we turn to the practically more interesting
three-dimensional (3D) space. The usefulness of the new model is illustrated
by a calculation of the amplitude coda in a random medium with a low density
of weak scatterers.

Let us consider a classical scalar field in 1D with spatially varying wave
velocity $c\left( x\right) .$ The 1D problem corresponds to a layered system
with planar sources and without lateral disorder, which is relevant as an
approximation for {\it e.g.} the seismics of rock sediments \cite{Kennet}.
The wave potential $\phi _{i}(x;\omega )$ at a given frequency $\omega $ is
then governed by the eigenvalue equation: 
\begin{equation}
\left\{ -\frac{\partial ^{2}}{\partial x^{2}}+V\left( x,\omega \right)
\right\} \phi _{i}(x;\omega )=E_{i}\left( \omega \right) \phi _{i}(x;\omega )
\end{equation}
introducing the frequency dependent ``scattering potential'' 
\begin{equation}
V\left( x,\omega \right) ={\frac{\omega ^{2}}{c_{0}^{2}}}\left( 1-\frac{%
c_{0}^{2}}{c\left( x\right) ^{2}}\right)
\end{equation}
which is ``attractive'' $\left( V<0\right) $ when the local velocity $%
c\left( x\right) $ is smaller than the reference velocity $c_{0}$ (as for
air bubbles in water) and ``repulsive'' $\left( V>0\right) $ when $c\left(
x\right) >c_{0}$ (as for metallic mercury in water). The physical quantity
of interest is the retarded Green function or point source propagator ($\eta 
$ is a positive infinitesimal): 
\begin{equation}
G(x,x^{\prime };\omega )=\sum_{i}{\frac{\phi _{i}(x;\omega )\phi
_{i}(x^{\prime };\omega )^{\ast }}{{\frac{\omega ^{2}}{c_{0}^{2}}}%
-E_{i}(\omega )+i\eta \text{sgn}(\omega )},}
\end{equation}
which is an observable for classical fields. A useful concept is the
scattering matrix which connects the amplitude of incoming and outgoing
amplitudes of a given scattering region \cite{Merz}. In 1D, the scattering
matrix has rank 2 and eigenvalues $S_{\ell },$ for the even channel $\ell =s$
and the odd channel $\ell =a.$

Let us consider a point model scattering potential\ $V_{\delta }\left(
x,\omega \right) \approx {\gamma }\left( \omega {/c_{0}}\right) ^{2}{\delta }%
\left( x\right) $ with a scattering \ strength parametrized by ${\gamma .\ }$%
A straightforward calculation gives: 
\begin{equation}
S_{s}^{\delta }\left( \omega \right) =-\frac{\omega +i\Gamma _{\delta }}{%
\omega -i\Gamma _{\delta }};\;S_{a}^{\delta }=1,
\end{equation}
where $\Gamma _{\delta }=2c_{0}/\gamma .$ The reflected amplitude from a
pulse at time $t^{\prime }=0$ and $x^{\prime }=L$ observed at time $t$ and
the same position $x=L$ can be obtained by contour integration, collecting
the pole at $\omega =i\Gamma _{\delta }$: 
\begin{gather}
G\left( L,L;t\right) =\int \frac{d\omega }{2\pi }\frac{c}{2i\omega }\frac{1}{%
2}\left( S_{s}-S_{a}\right) e^{i\omega 2L/c_{0}}e^{-i\omega t} \\
=\left. 
\begin{array}{c}
\frac{c_{0}}{2}e^{\Gamma _{\delta }\left( t-2L/c_{0}\right) }\Theta \left( t-%
\frac{2L}{c_{0}}\right) \\ 
-\frac{c_{0}}{2}e^{\Gamma _{\delta }\left( t-2L/c_{0}\right) }\Theta \left( 
\frac{2L}{c_{0}}-t\right)
\end{array}
\right\} \ \text{for}\ \left\{ 
\begin{array}{c}
\Gamma _{\delta }<0 \\ 
\Gamma _{\delta }>0
\end{array}
\right. ,
\end{gather}
where $\Theta $ is the step function. Clearly, the result is well behaved
when the scattering potential is attractive ($\Gamma _{\delta }<0),$ but
violates causality for a repulsive scatterer, since the reaction appears
before the action.

In order to shed light on this artifact, let us consider a finite square
well insertion of thickness $d$ and celerity (wave velocity) $c$ centered at
the origin and embedded into the infinite medium with celerity $c_{0}:$ 
\begin{equation}
V\left( x,\omega \right) =\frac{\omega ^{2}d}{c_{0}^{2}}\left( {1}-\frac{%
c_{0}^{2}}{c^{2}}\right) \frac{\Theta \left( d/2-x\right) -\Theta \left(
x-d/2\right) }{d}  \label{sw}
\end{equation}
In the limit of vanishing $d$ the last factor approaches the ${\delta }${%
-function. For a finite scattering amplitude in this limit }$\left|
1-c_{0}^{2}/c^{2}\right| $\ must scale like $d^{-1}$, which is possible when 
$c\ll c_{0}.$ When $c>c_{0},$ the scattering strength necessarily vanishes
when $d\rightarrow 0$. A repulsive point scatterer can only be realized by
an imaginary celerity in the insertion, which causes the non-causality.

Obviously, a better behaved model which retains the attractive features of
the point scatterer is necessary. We show now that a single pole
approximation of the scattering matrix \cite{Merz} satisfies these
requirements. The eigenvalues of the $S$-matrix for the potential (\ref{sw})
reads: 
\begin{equation}
S_{s/a}=\pm e^{-i\omega d/c_{0}}\frac{{\frak r}_{0}\left( 1-e^{2i\omega
d/c}\right) \mp \left( 1-{\frak r}_{0}^{2}\right) e^{2i\omega d/c}}{1-{\frak %
r}_{0}^{2}e^{2i\omega d/c}}
\end{equation}
where ${\frak r}_{0}=\left( c-c_{0}\right) /\left( c+c_{0}\right) $ is the
reflection coefficient of an isolated celerity discontinuity \cite
{Reflection}. The denominators of both $S_{s/a}$ vanish at 
\begin{equation}
\omega _{n}=n\frac{\pi c}{d}+i\frac{c}{d}\log \left| {\frak r}_{0}\right|
=n\Omega _{0}+i\Gamma .
\end{equation}
All resonances (or quasi-normal modes \cite{Alec}) are equally broadened by $%
\Gamma \ $and define the product representation of the scattering matrix
(with the proper global phase factor) \cite{Nico}$:$ 
\begin{equation}
S_{s/a}\left( \omega \right) =-e^{-i\omega d/c_{0}}\frac{\omega +i\Gamma }{%
\omega -i\Gamma }\prod_{n=1}^{\infty }\frac{\left( \omega +i\Gamma \right)
^{2}-\left( n\Omega _{0}\right) ^{2}}{\left( \omega -i\Gamma \right)
^{2}-\left( n\Omega _{0}\right) ^{2}}
\end{equation}
For the attractive (repulsive) scatterer the residues vanish when $n$ is odd
(even) for $\ell =s$ and $n$ is even (odd) for $\ell =a$. The poles are
always in the lower half of the complex frequency plane as required by
causality.

We can now introduce an approximation in which only the purely imaginary
pole is taken into account, which is formally justified for long time scales 
$\Delta t=t-2L/c_{0}\gg d/c$ and high reflectivities $1-{\frak r}_{0}^{2}\ll
1$. For an attractive scatterer 
\begin{equation}
S_{s}^{-}\approx -e^{-i\omega d/c_{0}}\frac{\omega +i\Gamma }{\omega
-i\Gamma };\;S_{a}^{-}\approx 1
\end{equation}
whereas for the repulsive scatterer the roles of the odd and even scattering
channels are reversed. The time-dependent reflection amplitude is now well
behaved: 
\begin{equation}
G^{\pm }\left( L,L;t\right) =\mp \Theta \left( t-\frac{2L}{c_{0}}\right)
\left| {\frak r}_{0}\right| ^{\frac{c}{d}\left( t-\frac{2L}{c_{0}}\right) +%
\frac{c}{c_{0}}}
\end{equation}

Let us now turn to the 3D problem. The $s$-wave scattering matrix of a point
scatterer at the origin $V\left( \vec{r},\omega \right) ={\gamma }\left(
\omega {/c_{0}}\right) ^{2}{\delta }\left( \vec{r}\right) $\ reads \cite
{Rossum}: 
\begin{equation}
S_{0}^{\delta }\left( \omega \right) =-\frac{\omega -i\frac{4\pi c}{\gamma }%
\left( \frac{c_{0}^{2}}{\omega ^{2}}-\frac{1}{k^{\ast 2}}\right) }{\omega -i%
\frac{4\pi c}{\gamma }\left( \frac{c_{0}^{2}}{\omega ^{2}}-\frac{1}{k^{\ast
2}}\right) },
\end{equation}
where $\left( k^{\ast }\right) ^{-1}=\sqrt{-\gamma k_{c}}$ and $k_{c}$ is a
high momentum cut-off which is necessary to regularize the point scatterer
model. Clearly $S_{0}^{\delta }$ is unitary for attractive scatterers only $%
\left( \gamma <0\right) \ $and does not have the correct analytical
properties for all $\gamma $. This do not disqualify point scatterers for
studies of low velocity insertions and monochromatic illumination, but they
are clearly unsuitable for pulsed, broad-band sources. We can find a remedy
along the lines sketched above for 1D using a spherical square-well
scatterer with diameter $d$ and velocity $c$. The higher-angular momentum
channels are of higher order in $\omega d/c_{0}$ and can therefore be
disregarded when $\Delta t\gg d/c_{0}$. The $s$-channel eigenvalue of the $S$%
-matrix read \cite{Merz}:

\begin{equation}
S_{0}\left( \omega \right) =-e^{-i\omega d/c_{0}}\left( 1+\frac{2i}{\frac{%
c_{0}}{c}\cot \frac{\omega d}{2c_{0}}-i}\right) .
\end{equation}
For repulsive scatterers its poles are at $\omega _{0,n}^{+}=2n\pi
c/d+i\Gamma $ with $\Gamma =\frac{c}{d}\log \left| {\frak r}_{0}\right| $.
Taking into account only the lowest, evanescent mode 
\begin{equation}
S_{0}^{+}\left( \omega \right) \approx -e^{-i\omega d/c_{0}}\frac{\omega
+i\Gamma }{\omega -i\Gamma }.  \label{rep}
\end{equation}
For attractive scatterers there is no purely imaginary pole since $\omega
_{0,n}^{-}=\left( 2n+1\right) \pi c/d+i\Gamma $ and the lowest approximation
involves two poles: 
\begin{equation}
S_{0}^{-}\left( \omega \right) \approx e^{-i\omega d/c_{0}}\frac{\left(
\omega +i\Gamma \right) ^{2}+\left( \frac{\pi c}{d}\right) ^{2}}{\left(
\omega -i\Gamma \right) ^{2}+\left( \frac{\pi c}{d}\right) ^{2}}  \label{att}
\end{equation}
The similarities and differences between 1D and 3D are notable. The exact
propagator in time 
\begin{eqnarray}
\Delta G_{\ell =0}\left( L,L;t\right) &=&-\frac{c_{0}}{8\pi L^{2}}\left[ 
{\frak r}_{0}\Theta \left( t-\frac{2L-d}{c_{0}}\right) \right.  \nonumber \\
&&\left. -\Theta \left( t-\frac{2L}{c_{0}}\right) \frac{4c_{0}c}{\left(
c+c_{0}\right) ^{2}}\sum_{n=0}^{\infty }{\frak r}_{0}^{n}\Theta \left( t-%
\frac{2L-d}{c_{0}}-\frac{\left( n+1\right) d}{c}\right) \right]
\label{exact3d}
\end{eqnarray}
is then approximated for $c>c_{0}$ as: 
\begin{eqnarray}
\Delta G_{\ell =0}^{+}(L,L;t) &\approx &-\frac{c_{0}}{8\pi L^{2}}\left[
\Theta \left( t-\frac{2L}{c_{0}}\right) +\Theta \left( t-\frac{2L-d}{c_{0}}%
\right) \left( 2{\frak r}_{0}^{\frac{c}{d}\left( t-\frac{2L-d}{c_{0}}\right)
}-1\right) \right]  \nonumber \\
&&\stackrel{d\rightarrow 0}{\approx }-\frac{c_{0}}{4\pi L^{2}}\Theta \left(
t-\frac{2L}{c_{0}}\right) {\frak r}_{0}^{\frac{c}{d}\left( t-\frac{2L}{c_{0}}%
\right) }  \label{2ndplus}
\end{eqnarray}
and $c<c_{0}$ as: 
\begin{eqnarray}
\Delta G_{\ell =0}^{-}(L,L;t) &\approx &-\frac{c_{0}}{8\pi L^{2}}\Theta
\left( t-\frac{2L}{c_{0}}\right) +\frac{c_{0}}{8\pi L^{2}}\Theta \left( t-%
\frac{2L-d}{c_{0}}\right)  \nonumber \\
&&\left\{ 1+\frac{4\log \left| {\frak r}_{0}\right| }{\pi }\left| {\frak r}%
_{0}\right| ^{\frac{c}{d}\left( t-\frac{2L-d}{c_{0}}\right) }\sin \left[ 
\frac{\pi c}{d}\left( t-\frac{2L-d}{c_{0}}\right) \right] \right\} \\
&&\stackrel{d\rightarrow 0}{\approx }\frac{c_{0}}{2\pi L^{2}}\Theta \left( t-%
\frac{2L}{c_{0}}\right) \frac{\log \left| {\frak r}_{0}\right| }{\pi }\left| 
{\frak r}_{0}\right| ^{\frac{c}{d}\left( t-\frac{2L}{c_{0}}\right) }\sin %
\left[ \frac{\pi c}{d}\left( t-\frac{2L}{c_{0}}\right) \right]
\label{2ndmin}
\end{eqnarray}
In Eqs. (\ref{2ndplus},\ref{2ndmin})\ the exponential phase factors in Eqs. (%
\ref{rep},\ref{att}) have\ been approximated by unity. In Figure 1 we
compare the exact reflection amplitudes (\ref{exact3d}) with the two
approximations. The envelope of the response is quantitatively well
represented by the lowest pole approximation. The subsequent neglect of the
phase shift is a rather crude approximation for a single scatterer, but
should be allowed in a many-scatterer configuration to be considered next.

Let us now turn to the response of a disordered medium. The approximate
scattering matrices derived above are directly related to the {\rm t}-matrix
eigenvalues ${\rm t}_{\ell }=c_{0}\left( S_{\ell }-1\right) /2i\omega ,$
which are proper partial sums in a perturbation theory of disordered systems
with non-overlapping scatterers \cite{Rossum}. In the limit of a low density 
$n$ of randomly distributed scatterers the interference between waves
multiply scattered by different sites may be disregarded. When all
scatterers are the same, the ensemble averaged Green function in wave vector 
$\vec{q}$ and frequency space reads (in 3D): 
\begin{equation}
\left\langle G(q;\omega )\right\rangle ^{-1}=\left( \frac{\omega }{c_{0}}%
\right) ^{2}-q^{2}-n{\rm t}_{0}
\end{equation}
For repulsive scatterers we employ the approximation (\ref{2ndplus}).
Transformation into time and position space gives: 
\begin{equation}
\left\langle G(r;t)\right\rangle =-\frac{\Theta \left( t-r/c_{0}\right) }{%
2\pi r}\int_{0}^{\infty }\frac{d\omega }{2\pi }\cos \left( \omega t-r\left| 
\mathop{\rm Re}%
\kappa \right| \right) e^{-\left| 
\mathop{\rm Im}%
\kappa \right| r}
\end{equation}
where $\kappa =\sqrt{\left( \omega /c_{0}\right) ^{2}-n{\rm t}_{0}}$ is the
``renormalized'' wave vector. In the weak scattering limit, which holds for
time scales $\Delta t\gg \left| \Gamma \right| ^{-1}:$ 
\begin{equation}
\left\langle G(r;t)\right\rangle =-\frac{e^{-\frac{2\pi nc_{0}^{3}}{\Gamma
^{2}}t}}{4\pi r}\left[ \delta \left( t-r/c_{0}\right) +\Theta \left(
t-r/c_{0}\right) C(r;t)\right] ,  \label{coh}
\end{equation}
where the delayed signal or amplitude ``coda'' reads 
\begin{equation}
C(r;t)=\frac{1}{\pi t}\int_{0}^{\infty }dx\left[ \cos \left( x-\frac{r}{%
c_{0}t}\sqrt{x^{2}-\frac{4\pi nc_{0}^{3}t^{2}}{\Gamma }}\right) -\cos \left(
x-\frac{r}{c_{0}t}x\right) \right] .  \label{coda}
\end{equation}
This integral can be solved analytically for the first arrival: 
\begin{equation}
C(c_{0}t;t)=\frac{1}{2\pi t}\left[ \sin \sqrt{\frac{4\pi nc_{0}^{3}}{\left|
\Gamma \right| }}t-\sqrt{\frac{4\pi nc_{0}^{3}}{\left| \Gamma \right| }}%
t\cos \sqrt{\frac{4\pi nc_{0}^{3}}{\left| \Gamma \right| }}t-\frac{4\pi
nc_{0}^{3}}{\left| \Gamma \right| }t^{2}%
\mathop{\rm Si}%
\sqrt{\frac{4\pi nc_{0}^{3}}{\left| \Gamma \right| }}t\right] .
\label{analytic}
\end{equation}
The amplitude is (Lambert-Behr) exponentially damped by the imaginary part
of the t-matrix. The time-delayed signal plotted in Fig. 2 originates from
multiple scattering effects on the real part of the refractive index which
is equivalent with a reduced effective velocity. It depends
characteristically on the microscopic parameters of the systems, {\it viz}.
the density, size, and the wave velocity (contrast) of the scatterers. Note
that the long-lived incoherent fluctuations which arrive at the detector
after diffusion (the intensity coda) contribute only to the intensity $%
\left\langle G^{2}(r;t)\right\rangle ,$ which is not discussed here.

In conclusion, we have revealed the inappropriateness of point scatterers
for time-dependent classical wave scattering. An alternative model in terms
of a lowest-order pole expansion is simple and well behaved. The propagator
of a homogeneously disordered medium is calculated with the alternative
model. A delayed signal or ``coda'' is found, which depends on the
microscopic parameters of the scatterers. These results support speculations
that the coda contains important information about a disordered medium that
might be relevant for imaging applications.

We thank Alec Maassen van den Brink and Yaroslav Blanter for critically
reading the manuscript and Yuli Nazarov for help in finding Eq. (\ref
{analytic}). This work is part of the research program of the ``Stichting
Technische Wetenschappen'' (STW) and the ``Stichting Fundamenteel Onderzoek
der Materie (FOM)''. G.B. acknowledges support by the NEDO program NTDP-98.

\begin{figure}[tbp]
\caption{Time-dependence of the amplitude originating from a
pulsed source and reflected by a spherical square-well potential in the $s$-channel; (a) high velocity (b)
low velocity scatterer of diameter $d$. $\Delta t=0$ corresponds to the
nominal arrival time. The inserts show the velocity profile in units of $c_0$ and the radius of the scatterer. Full curves denote
the exact results, dashed curves correspond to the lowest pole approximation
and in the dotted curves the phase shift has been disregarded.}
\end{figure}
%

%
\begin{figure}[tbp]
\caption{Time delayed signal in the coherently propagated amplitude $C(r=l,t)$
according to Eqs. (\ref{coh},\ref{coda}) in a disorderd medium with different scattering
parameters $\Gamma \protect\tau $=5 (full curve), 10 (dashed), 15 (dotted), where $l=c_{0}\protect\tau
=\Gamma ^{2}/(4\protect\pi nc_{0}^{2})$ is the mean free path, 
$n$ the density of scatterers and $\Gamma 
$ the imaginary part of the single-scatterer resonance frequency.}
\end{figure}

%


\begin{references}
\bibitem{general}  P. Sebbah (ed.), {\it Waves and Imaging through Complex
Media} (Kluwer, Dordrecht, 2001) and references therein.

\bibitem{Wiersma}  D.S. Wiersma, P. Bartolini, A. Lagendijk, and R. Rhigini,
Nature {\bf 390}, 671 (1997); J. de Rosny, A. Tourin, and M. Fink, Phys.
Rev. Lett. {\bf 84}, 1693 (2000).

\bibitem{Lee}  P.A. Lee and T.V. Ramakrishnan, {\it Anderson Localisation},
Rev. Mod. Phys. {\bf 57}, 287 (1985).

\bibitem{Abrikosov}  A.A. Abrikosov, L.P. Gorkov, and I.E. Dzyaloszinski, 
{\it Methods of Quantum Field Theory in Statistical Physics} (Prentice Hall,
New Jersey, 1963).

\bibitem{Rossum}  M.C.W. van Rossum and Th.M. Nieuwenhuizen, {\it Multiple
scattering of classical waves: microscopy, mesoscopy, and diffusion}, Rev.
Mod. Phys. {\bf 71}, 313 (1999).

\bibitem{Vries}  P. de Vries, D.V. van Coevorden and A. Lagendijk, {\it %
Point scatterers for classical waves}, Rev. Mod. Phys. {\bf 70}, 447 (1998).

\bibitem{Snieder}  M. Roth, G. Mueller, and R. Snieder, Geophys. J. Int. 
{\bf 115}, 552 (1993); J. Tworzydlo and C.W.J. Beenakker, Phys. Rev. Lett. 
{\bf 85}, 674 (2000).

\bibitem{Coda}  H. Sato and M.C. Fehler, {\it Seismic Wave Propagation and
Scattering in the Heterogeneous Earth} (Springer, New York, 1998).

\bibitem{White}  B. White, P. Sheng, Z.Q. Zhang, and G. Papanicolaou, Phys.
Rev. Lett. {\bf 59}, 1918 (1987).

\bibitem{Titov}  M. Titov and C.W.J. Beenakker, Phys. Rev. Lett. {\bf 85},
3388 (2000).

\bibitem{Beenakker}  C.W.J. Beenakker, Rev. Mod. Phys. {\bf 69}, 731 (1997).

\bibitem{Kennet}  B.L.N. Kennett, {\it Seismic Wave Propagation in
Stratified Media}, Cambridge University Press, (1983).

\bibitem{Merz}  E. Merzbacher,``Quantum Mechanics'' (Wiley, New York, 1961).

\bibitem{Reflection}  Note that mass density differences $\rho -\rho _{0}$
can be easily incorporated by ${\frak r}_{0}=\left( \rho c-\rho
_{0}c_{0}\right) /\left( \rho c+\rho _{0}c_{0}\right) $

\bibitem{Nico}  N.G. van Kampen, Phys \ Rev. {\bf 91}, 1267 (1953); H.M.
Nussenzweig, {\it Causality and Dispersion Relations} (Academic, New York,
1972).

\bibitem{Alec}  E.S.C. Ching, P.T. Leung, A. Maassen van den Brink, W.M.
Suen, S.S. Tong, K. Young, {\it Quasinormal-mode expansion for waves in open
systems}, Rev. Mod. Phys. {\bf 70}, 1545 (1998).
\end{references}
\end{document}